\documentclass[aps,prl,twocolumn
]{revtex4-2}

\usepackage{natbib}
\usepackage{amsmath}
\usepackage{amsfonts}
\usepackage[makeroom]{cancel}
\usepackage{graphicx}
\usepackage{mathrsfs}
\usepackage{euscript}
\usepackage{color}
\usepackage{dsfont}
\usepackage{url}
\usepackage[colorlinks,citecolor=blue]{hyperref}
\usepackage{hyperref}
\hypersetup{
    colorlinks=true,
    linkcolor=blue,
    filecolor=blue,
    urlcolor=blue,
}

\begin{document}


\title{Fractionalized topological $d+id$ 
superconductivity in the Yao-Lee-Kondo model}


\author{Chengzhi Tang}
\affiliation{Institute for Advanced Study, Tsinghua University, Beijing 100084, China}
\author{Hong Yao}
\email{yaohong@tsinghua.edu.cn}
\affiliation{Institute for Advanced Study, Tsinghua University, Beijing 100084, China}


\date{\today}

\begin{abstract}

A conclusive experimental realization of 2D chiral topological superconductivity remains elusive. Here we present a theoretical demonstration that a topological $d+id$ fractionalized superconducting phase (SC*) can emerge in the weak-coupling limit of a Kondo lattice model, where conduction electrons interact with a Yao-Lee spin liquid on the honeycomb lattice (the Yao-Lee-Kondo model). Using a renormalization-group analysis, we show that exchanging Majorana spinons from the Yao-Lee spin liquid generates effective interactions among the conduction electrons and drives a Cooper instability even for arbitrarily weak Kondo coupling. We further find that the induced leading inter-orbital antiferromagnetic interaction selects
topological $d+id$ spin-singlet pairing with Chern number $C=\pm 2$. Meanwhile, the Majorana fermions in the Yao-Lee spin liquid remain gapless and deconfined in this regime, so the resulting state is a fractionalized topological $d+id$ superconductor (SC*). For sufficiently strong Kondo coupling, the system instead enters a heavy Fermi liquid phase with fractionalization (HFL*).
\end{abstract}


\maketitle


\textit{Introduction.}---The Kondo lattice system constitutes a canonical setting for studying strongly correlated electrons \cite{Kondo_lattice1,Kondo_lattice2,Kondo_lattice3,Kondo_lattice4,Kondo_lattice5,Kondo_lattice_review1,Kondo_lattice_review2,Kondo_lattice_review3,jiao2015fermi}. Experiments on Kondo lattice compounds such as $\mathrm{CeCu_2Si_2}$ and $\mathrm{UPt_3}$ have revealed various intriguing features, including 
emergent heavy fermion behavior \cite{heavy_fermion,heavy_fermion_dis} and unconventional superconductivity (SC) \cite{uSC_CeCuSi,uSC_Ube13,uSC_UPt3,uSC_UTe2,uSC_CeCoIn5, yuan2003, pang2018fully}. 
On one hand, the pronounced Kondo screening effect drives the system into a heavy Fermi liquid state, characterized by a strongly renormalized electronic density of states in the vicinity of the Fermi level \cite{heavy_fermion,heavy_fermion_dis,HFL1,HFL2,HFL3,HFL4,HFL5,HFL6,HFL7,HFL8,HFL9}. In contrast, magnetic Ruderman-Kittel-Kasuya-Yosida (RKKY) exchange interactions compete with the Kondo effect and tend to stabilize the antiferromagnetic ordering of the local moments \cite{RKKY1,RKKY2,RKKY3,AFM_evidence1,AFM_evidence2}. 
Moreover, $d$-wave SC has been experimentally observed near antiferromagnetic quantum critical points \cite{d-wave_evidence1,d-wave_evidence2,d-wave_evidence3,d-wave_evidence4,d-wave_evidence5} in certain heavy fermion materials. The pairing of heavy quasiparticles is widely believed to be driven by magnetic quantum critical fluctuations associated with these instabilities \cite{spin_glue1,spin_glue2,spin_glue3,spin_glue4,spin_glue5}.

\begin{figure}
\includegraphics[width=0.9\linewidth]{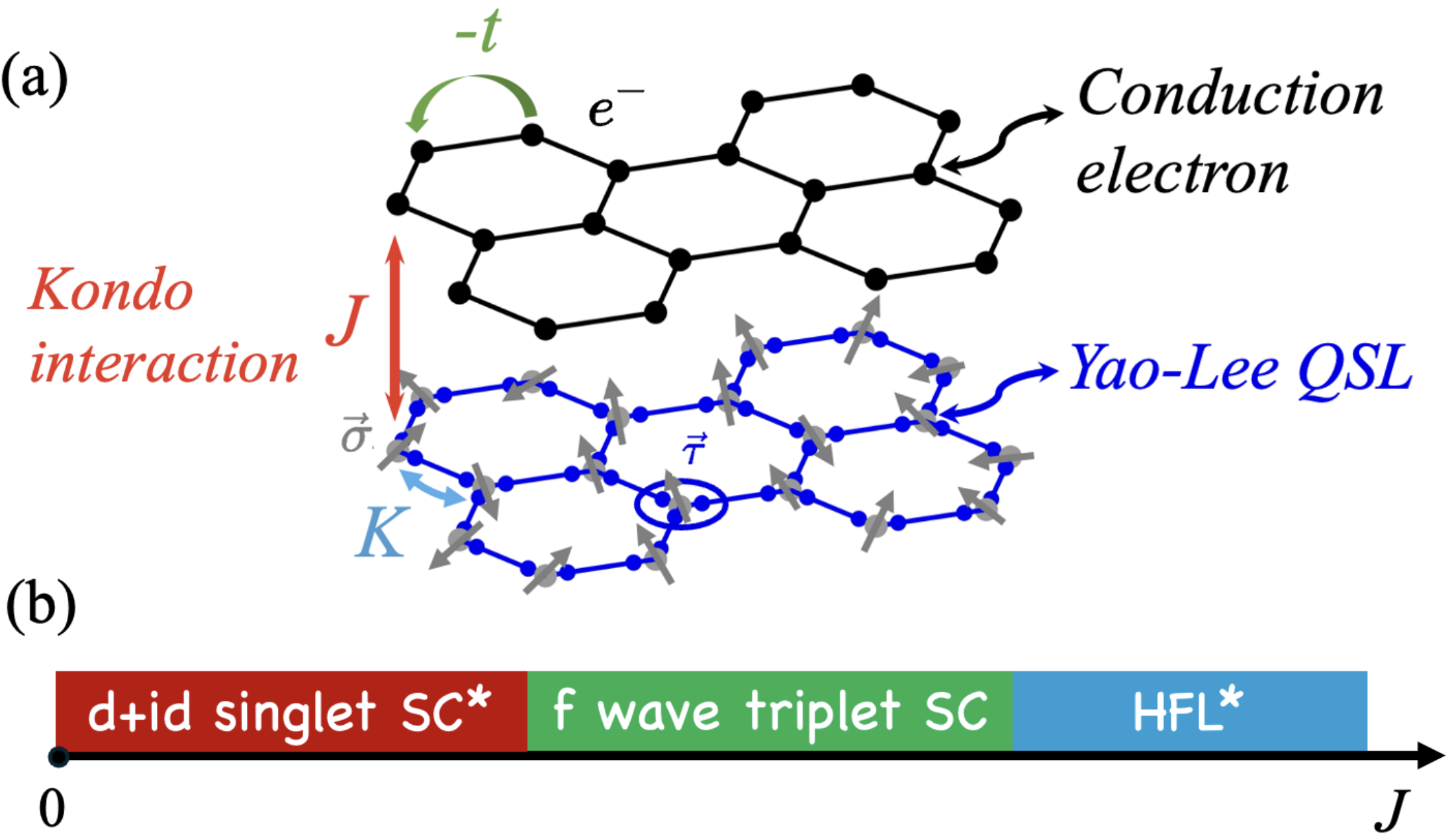}
\caption{\label{fig:Model} (a) A schematic representation of the Yao-Lee-Kondo model on the honeycomb lattice. The first (black) layer represents the conduction electrons, which are Kondo coupled to the Yao-Lee spins residing in the second layer. In the Yao-Lee model, each localized electron carries both spin and orbital indices.
(b) Quantum phase diagram of the Yao-Lee-Kondo model. In the weak-coupling regime, the system realizes a topological $d+id$ spin-singlet superconductivity coexisting with an underlying Yao-Lee spin liquid (SC*). 
At sufficiently strong coupling, a heavy Fermi-liquid phase arises from hybridization between the conduction electrons and Yao-Lee spins while the Yao-Lee orbital moments simultaneously form an underlying a quantum orbital liquid, realizing a fractionalized heavy Fermi liquid (HFL*).}
\end{figure}

One natural question that has been asked is whether—and by what mechanism—superconductivity can emerge in a Kondo lattice where the localized moments form a quantum spin liquid; a quantum spin liquid lacks any conventional long-range order but supports deconfined spinon excitations \cite{QSL-anderson1,QSL-anderson2,QSL-fm,QSL-RK,QSL-Kivelson,QSL-MR,QSL-MRN,QSL-PALee,QSL-TK,KitaevQSL-Kee,QSL1,QSL2}. For sufficiently strong Kondo coupling, it has been shown 
that deconfined spinons in quantum spin liquids can act as an effective “pairing glue”, mediating pairing between conduction electrons \cite{fractionalized_FL}. 
More recently, two types of 2D Kondo lattice models coupled to exactly-solvable $Z_2$ spin liquids have been investigated in this context: the Kitaev-Kondo (KK) model \cite{Kitaev-Kondo1,Kitaev-Kondo3} where the conduction electrons couple with the Kitaev spin liquid \cite{kitaev2006anyons} on the honeycomb lattice, and the Yao-Lee-Kondo (YLK) model \cite{Kitaev-Kondo2} where the conduction electrons couple with the Yao-Lee spin liquid on the honeycomb lattice \cite{Yao-Lee}. A roadmap for realizing the Yao-Lee model in quantum materials was recently proposed \cite{Yao-Lee_Roadmap}. In contrast to the Coleman-Panigrahi-Tsvelik (CPT) model \cite{CPT-model,FFLO-CPT,QSL_octagon_lattice} on the 3D hyper-octagon lattice where the Fermi surface of conduction electrons and that of spinons are perfectly nested, the spinons in the Kitaev spin liquid or Yao-Lee spin liquid on the honeycomb lattice feature massless Dirac dispersion with a vanishing density of states at the Dirac point. Whether SC emerges in a Kondo lattice where conduction electrons are weakly coupled with a quantum spin liquid containing massless Dirac spinons remains elusive. In the limit of weak Kondo coupling, one might expect that the conduction electrons form a stable Fermi liquid co-existing with the spin liquid (namely FL* \cite{fractionalized_FL}), due to the vanishing density of states of spinons. 

However, in this paper, we shall demonstrate that the Fermi surface of the putative FL* state is generically unstable, regardless of how weak the Kondo coupling is, and that fractionalized superconductivity SC* arises in the ground state. To illustrate this, we focus on the Yao-Lee-Kondo model on the honeycomb lattice, illustrated in Fig.~\ref{fig:Model}(a), where conduction electrons are coupled with a Yao-Lee spin liquid on the honeycomb lattice \cite{Kitaev-Kondo2}. 

Due to the computational complexity and the fermion sign problem \cite{ZXLi-PRB2015,ZXLi-PRL2016,Sign_problem}, it is challenging to study the Yao-Lee-Kondo lattice using numerical techniques such as DMRG or quantum Monte Carlo for systems of large sizes \cite{DMRG1,DMRG2,QMC_review,Sign_problem}. 
However, owing to the unique feature of intact $Z_2$ gauge structures in the Kondo coupling of the Yao-Lee-Kondo model, we can perform a perturbatively exact renormalization-group (RG) analysis \cite{shankar1994renormalization,RG-Kivelson} and thereby fully characterize the weak-coupling limit of the Yao-Lee-Kondo model. We find that the induced antiferromagnetic interaction between electrons, mediated by exchanging Majorana spinons, drives a Cooper instability of the Fermi surface and stabilizes a topological $d+id$ spin-singlet superconducting state with Chern number $C=\pm2$. The Yao-Lee Majorana sector remains gapless, preserving the fractionalized excitations of the original Yao-Lee model in the fractionalized SC* phase. 

Building on our RG analysis, we propose the phase diagram of the Yao-Lee-Kondo model shown in Fig. \ref{fig:Model}(b). Recent mean-field calculations \cite{Kitaev-Kondo2} and numerical studies of the $Z_2$-gauged XY model \cite{gauged-XY} indicate that, at intermediate Kondo coupling $J$, the system enters a superconducting phase with $f$-wave spin-triplet pairing. Unlike the $d+id$ SC* phase, the $f$-wave spin-triplet SC phase does not host a topological order because of the condensation of hybridization between Majorana fermions and conduction electrons \cite{Kitaev-Kondo2}. In the strong-coupling regime, the system enters a fractionalized heavy Fermi liquid (HFL*) phase \cite{CPT_breakdown,CPT_phase_diagram,vison_gap,Ancilla-layer}, where the conduction electrons hybridize with the Yao-Lee spins to form the HFL, while the Yao-Lee orbital degrees of freedom realize an underlying Kitaev orbital liquid.

\textit{Model.}---The Hamiltonian of the Yao-Lee-Kondo lattice, in which the conduction electrons are coupled with a Yao-Lee spin liquid on the honeycomb lattice, has three parts $ \hat H_\textit{YLK} = \hat H_\textit{YL} + \hat H_c + \hat H_K$:
\begin{equation}
\begin{aligned}
    &\hat H_\textit{YL} =  K\sum_{\left \langle ij\right \rangle}\tau^{\lambda_{ij}}_i \tau^{\lambda_{ij}}_j({\vec{S}}_i\cdot{\vec{S}}_j),\\
    &\hat{H}_c = -t\sum_{\left \langle ij\right \rangle,\sigma}(c^\dagger_{i\sigma}c_{j\sigma} + \mathrm{H.c.}) -\mu\sum_{i\sigma}c^{\dagger}_{i\sigma}c_{i\sigma},\\
    & \hat H_K = J\sum_{i}\vec S_i\cdot  c^{\dagger}_i\vec{\sigma}c_i,
\end{aligned}
\end{equation}
$\hat H_\textit{YL}$ and $\hat H_c$ are the Hamiltonians of the Yao-Lee spin-orbit model and the conduction electrons, respectively. $\tau/\sigma$ are Pauli operators for the orbital and spin degrees of freedom of localized electrons; $\lambda_{ij}={x,y,z}$ is the bond direction of link $\left \langle ij\right \rangle$; $\vec S_i=\frac{1}{2}\vec\sigma_i$ are spin-1/2 operators; $\tau_i^\lambda$ are Pauli operators with orbital indices. 
Here, we set the Yao-Lee model in the isotropic limit with coupling $K$ and consider conduction electrons with a finite Fermi surface by setting $\mu>0$. 
The Yao-Lee model can be fermionized and exactly solved by employing the Majorana representation \cite{Yao-Lee}: $\sigma^{\alpha}_i=-\frac{i}{2}\epsilon^{\alpha\beta\gamma}\chi^{\beta}_i\chi^{\gamma}_i$ and $\tau^{\alpha}_i=-\frac{i}{2}i\epsilon^{\alpha\beta\gamma}d^{\beta}_id^{\gamma}_i$, with the local constraint $D_i=-id^{x}_id^{y}_{i}d^{z}_i\chi^{x}_i\chi^{y}_i\chi^{z}_i=1$ and Majorana fermion anti-commutation relations $\{\chi^{\alpha}_i,\chi^{\beta}_j\}=2\delta_{ij}\delta_{\alpha\beta}$ and $\{d^{\alpha}_i,d^{\beta}_j\}=2\delta_{ij}\delta_{\alpha\beta}$ \cite{Yao-Lee}. Under this representation, $H_\textit{YL}$ can be written in quadratic form:
\begin{equation}
     \hat H_\textit{YL} =\frac{K}{4}\sum_{\left \langle ij\right \rangle}  \hat u_{ij}(i\vec{\chi}_i\cdot\vec{\chi}_j).
\end{equation}
The $Z_2$ gauge field $\hat u_{ij}=-id^{\lambda_{ij}}_id^{\lambda_{ij}}_j=\pm1$. According to the Lieb's theorem, the gauge field $\hat u_{ij}$ takes a zero flux configuration in ground state. Then the Kondo interaction couples the spin part of the Yao-Lee model to the conduction electron spin in the form of Heisenberg interaction. In Majorana representation, the Kondo interaction corresponds to the process of electrons being scattered by pairs of Majorana spinons:
\begin{equation}
    \hat H_K =-J/2\sum_{i,a<b}  \chi^a_i  \chi^b_i \cdot c_i^\dagger\sigma^a\sigma^bc_i.
\end{equation}
Note that the Kondo interaction preserves the original $Z_2$ gauge structure since $[\hat H_K,\hat u_{ij}]=0$. Due to the spin $SU(2)$ rotational symmetry of the Yao-Lee model, the Yao-Lee-Kondo model with both localized spins and conduction electrons also respects the spin $SU(2)$ symmetry. 

We shall perform the RG calculation at the weak coupling limit $J\ll K,t$ and explore the Cooper instability of the Fermi surface of conduction electrons. 
The RG calculation proceeds in two steps: (i) Integrate out fast modes and obtain the effective action in terms of slow modes; 
(ii) Derive the RG equations. 

\textit{Low-energy effective action. 
}---We first obtain the effective $S_0$ for free conduction electrons and Majorana spinons with a momentum cutoff $\Lambda$. We then write down the action of the bare Kondo interaction $S_K$ in the low-energy regime. The action of free conduction electrons and Majoranas can each be decomposed into two valleys, labeled by $l=+/-$, corresponding to the $K/K^{\prime}$ points, respectively. We obtain $S=S_0+S_K$ with 
\begin{equation}
    \begin{aligned}
    S_0 =  \int_{\Lambda} &\mathrm{d}q \  \bar{\chi}^{a(l)}_{q}\frac{1}{2}[i\omega-v^{\chi}_F(q_x\eta_x+\mathrm{sgn}(l)\cdot q_y\eta_y )]\chi^{a(l)}_{q}\\
    &+\int_{\Lambda} \mathrm{d}k \ \bar\psi^{(l)}_{k}(i\omega-v^{c}_{F} \tilde{ k})\psi^{(l)}_{k},\\
    S_{K}=\int_{\Lambda}&\mathrm{d}k\mathrm{d}q_1\mathrm{d}q_2\ \sum_{a<b}\frac{J_\alpha^{l_3,l_4}(\theta_\mathbf k,\theta_{\mathbf k^*})}{2}\ \Theta(\Lambda-|\tilde k^{*}|)\\&\delta_{l_1+l_2+l_4-l_3}\ \chi^{a(l_1)}_{q_1,\alpha}\ \chi^{b(l_2)}_{q_2,\alpha}\cdot\bar \psi^{(l_3)}_{k^*}\sigma^{a}\sigma^{b}\psi^{(l_4)}_{k},
\end{aligned}
\label{S_bare}
\end{equation}
where $\psi_k$ represents the Grassmann number of the upper band basis ($\mu>0$) of conduction electrons with the spin index included, $\chi^{a(l)}_{q}$ is the two component Grassmann number including sublattice $A/B$ and $\eta_\alpha$ is the Pauli matrix in sublattice subspace.  $\tilde{k}=|\mathbf k|-K_F$ and $\mathbf{q}=(q_x,q_y)$, and the summation over repeated indices is assumed. The following notation is used: $k=(i\omega,\mathbf k)$ and the spatial momentum is measured against the corresponding Dirac point.  The momentum conservation gives $k^*=k+q_1+q_2$ and the constraint $\delta_{l_1+l_2+l_4-l_3}$. $J_\alpha^{l_3,l_4}(\theta_\mathbf k,\theta_{\mathbf k^*})$ denotes the Kondo coupling amplitude projected onto the $\psi$ band basis, whose explicit derivation is presented in the Supplemental Material. The Fermi velocity of conduction electron and Majorana band are $v^{c}_{F}=\frac{3}{2}t$, $v^{\chi}_{F}=\frac{3}{8}K$ respectively. The integral regime of momentum is restricted to the $2\Lambda$ shell. We point out that momentum $\mathbf k$ and $\mathbf q$ have different rescaling rules in the RG procedure: 
\begin{equation}
    \tilde{k}^{\prime}=b\tilde k\ \ \text{and} \ \ \mathbf q^{\prime}=b\mathbf q. 
\end{equation}
 
\begin{figure}
\includegraphics[width=0.9\linewidth]{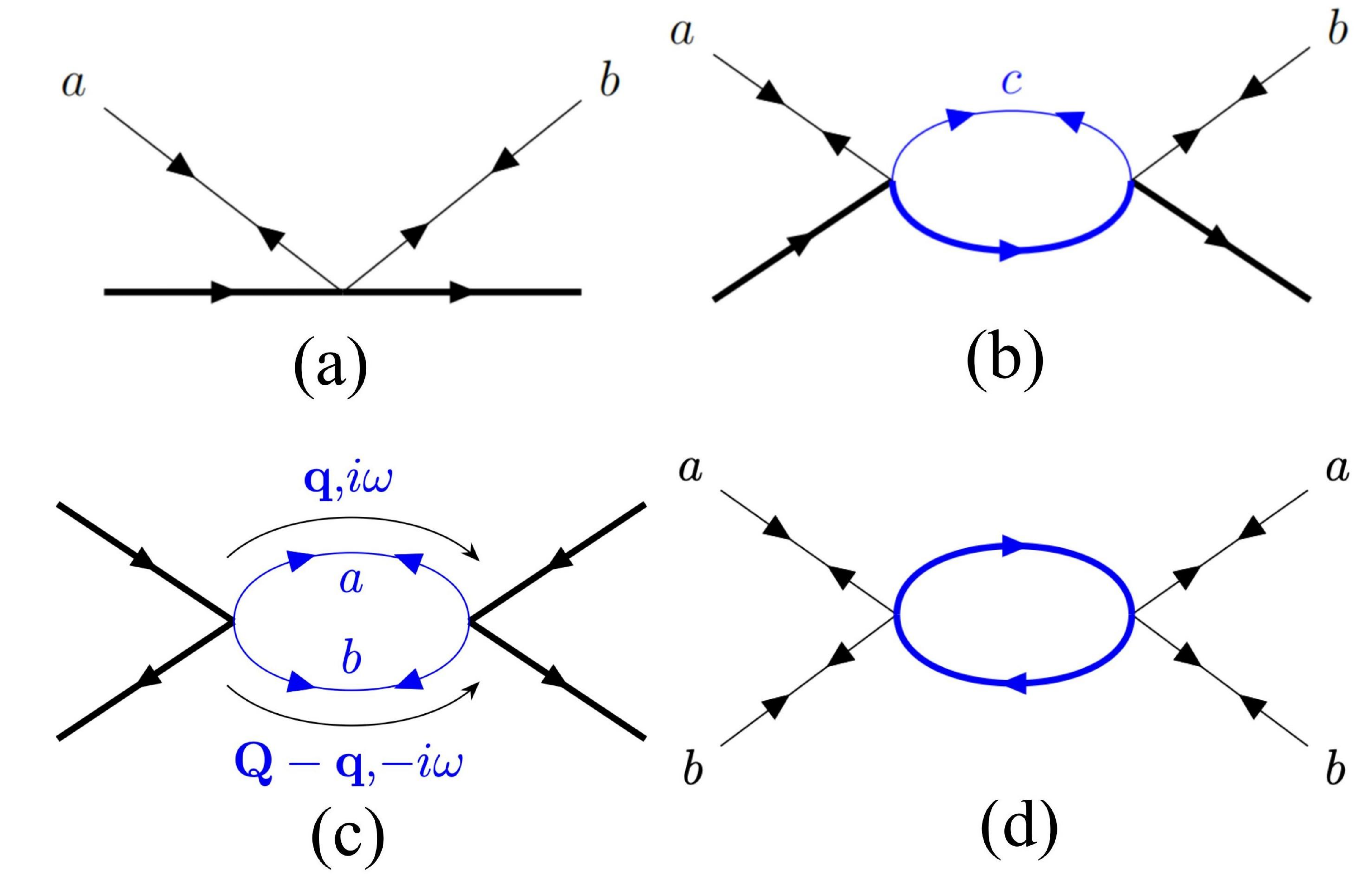}
\caption{\label{fig: Feynman diagrams} 
Panel (a) represents the bare Kondo interaction. Panels (b), (c), and (d) show the induced Kondo interaction, e-e interaction, and $\chi$-$\chi$ interaction at the one-loop level, respectively. The following conventions are used in the Feynman diagrams: (i) electron propagators are denoted by thick lines with a single arrow; (ii) Majorana propagators are denoted by thin lines with double arrows; and (iii) fast modes to be integrated out are indicated by blue lines and the slow modes are indicated by black lines. }
\end{figure}

Instead of computing the $S^{\mathrm{eff}}_{\mathrm{int}}$ exactly, we focus on the potentially relevant terms that arise after integrating out the fast modes and present the lowest-order Feynman diagrams corresponding to the different types of interactions. The quartic terms of $S^{\mathrm{eff}}_{\mathrm{int}}$ can be divided into three groups: (i) Kondo interaction; (ii) Interaction between conduction electrons $\psi$; (iii) Interaction between Majorana fermions $\chi$. The bare Kondo interaction and induced interaction of (i)(ii)(iii) are shown in Fig.~\ref{fig: Feynman diagrams}(a)(b)(c)(d) respectively. In the next section, we will show that interactions (i) and (iii) are irrelevant; therefore, we will concentrate here on the effective e–e interaction.

After integrating out the fast mode, the bare Kondo interaction also induces effective e-e interaction by exchanging a pair of Majorana spinons, as shown in Fig.\ref{fig: Feynman diagrams}(b). Here, $\mathbf{Q}$ denotes the momentum transfer in the process of electron scattering. The interaction $\delta S^\mathrm{eff}_{\mathrm{e-e}}$ can be decomposed into two parts: $\delta S^{(1)}_{\mathrm{e-e}}$ and $\delta S^{(2)}_{\mathrm{e-e}}$, corresponding to the contributions of $\mathbf Q\approx0$ and $\mathbf Q\approx\pm2\mathbf K$ respectively:
\begin{equation}
\begin{aligned}
      \delta S^{(1)}_{\mathrm{e}-\mathrm{e}} = \frac{V_1(\Lambda)J^2}{16K}&\int_{\Lambda} \mathrm dk\mathrm dk^\prime \ [1-\cos(\theta_\mathbf k-\theta_{\mathbf k^{\prime}})]\ \\ &\bar\psi^{(l)}_k\vec{\sigma}\psi^{(l)}_{k^\prime}\cdot\bar\psi^{(\bar l)}_{-k}\vec\sigma\psi^{(\bar l)}_{-k^{\prime}},
    \label{S1}  
\end{aligned}
\end{equation}
\vspace{-\abovedisplayskip}
\begin{equation}
    \delta S^{(2)}_\mathrm{e-e}=\frac{V_2(\Lambda)J^2}{16K}\int_{\Lambda} \mathrm{d}k\mathrm{d}k^{\prime}\ \bar\psi^{(l)}_k\vec{\sigma}\psi^{(\bar l)}_{k^{\prime}}\cdot\bar\psi^{(\bar l)}_{-k}\vec\sigma\psi^{(l)}_{-k^{\prime}}.
    \label{S2}
\end{equation}
 The interactions are written out in the Cooper channel. Dimensionless constants $V_1,V_2$ increase as the cutoff $\Lambda$ is reduced and converge to $V_1(0) \approx2.17, \ V_2(0)\approx1.80$. Here, $\theta_\mathbf k$ denotes the direction of $\mathbf{k}$ and $\bar l$ labels the valley opposite to $l$. The calculation details of $\delta S^{\mathrm {eff}}_{\mathrm{e-e}}$ are shown in the Supplemental Material. The induced e-e interaction will be analyzed by the RG approach in the next section.

\textit{RG analysis.}---We first focus on the tree level RG analysis of the Kondo interaction. After projecting onto the $\psi^{(l)}_k$ basis, the coupling strength takes the general form $J(k,q,q^{\prime},\Lambda)$. Within the RG framework, the $\tilde k,q,q^{\prime}$ dependence can be neglected:
\begin{equation}
\begin{aligned}
    S_{K}=\int_{\Lambda}\mathrm{d}k&\mathrm{d}q_1\mathrm{d}q_2\ \sum_{a<b}J^{l,l^\prime}_{\alpha}(\theta_\mathbf k,\Lambda)\ \Theta(\Lambda-|\tilde k^{*}|)\\&\chi^{a(l)}_{q_1,\alpha}\ \chi^{b(\bar l)}_{q_2,\alpha}\cdot\bar \psi^{(l^\prime)}_{k^*}\sigma^{a}\sigma^{b}\psi^{(l^{\prime})}_{k},
\end{aligned}
\end{equation}
where $\tilde k^{*}= |\mathbf k+\mathbf q_1+\mathbf q_2|-K_F$. For simplification, we redefine $\mathbf Q=\mathbf q_1+\mathbf q_2$ and $\mathbf Q^{\prime}=b\mathbf Q$ after rescaling. In the above equation, only the intra-valley scattering terms are shown as an example. After rescaling, the action is transformed to:
\begin{equation}
\begin{aligned}
    S_{K}=b^{-r}\int_{\Lambda}\mathrm{d}k^{\prime}&\mathrm{d}q_1^{\prime}\mathrm{d}q_2^{\prime}\ \sum_{a<b}J^{l,l^{\prime}}_\alpha(\theta_\mathbf {k^{\prime}},\Lambda)\ \Theta(b^{-1}\Lambda-|\tilde {k}^{*}|)\\ &\chi^{a(l)}_{q_1^{\prime},\alpha}\ \chi^{b(\bar l)}_{q_2^{\prime},\alpha}\cdot\bar \psi^{(l^\prime)}_{{k^*}^{\prime}}\sigma^{a}\sigma^{b}\psi^{(l^{\prime})}_{k^{\prime}},
\end{aligned}
\end{equation}
 where $r=[\mathrm dk]+2[\mathrm dq]+2[\psi]+2[\chi]$ and $[A]$ denotes the scaling dimension of $A$. The scaling dimensions of $\psi$ and $\chi$ are obtained by requiring $S_0$ invariant after rescaling \cite{shankar1994renormalization}:
\begin{equation} 
    [\mathrm dk]=2\ \ \  \ [\mathrm dq]=3\ \ \ \ [\psi]=-3/2 \ \ \ \ [\chi]=-2\ \ \ \ 
\end{equation}
From the above analysis, we obtain $r=1$ . To ensure that the RG procedure is well defined, it is also necessary that the step function $\Theta$ retains its original form under rescaling. Assuming $\Lambda/K_F\ll1$, the linear approximation about $\tilde k$ and $\mathbf Q$ can be adopted: $|\tilde {k^{*}}|\approx \tilde k +\mathbf{|Q|}\cos\theta_{\mathbf{k,Q}}$, where $\theta_{\mathbf{k,Q}}$ denotes the angle between momentum $\mathbf k$ and $\mathbf Q$. Although $K_F$ is not rescaled under RG operation, it drops out from the $\Theta$ function after the linear approximation is applied:
\begin{equation}
\begin{aligned}
    \Theta(b^{-1}\Lambda-|\tilde k^{*}|)
    &\approx \Theta[b^{-1}(\Lambda-\tilde k^{\prime}-|\mathbf Q^{\prime}|\cos\theta_{\mathbf{k,Q}})]\\
    &\approx\Theta(\Lambda-|\tilde {k^{*}}^{\prime}|).
\end{aligned}
\end{equation}
Now the $\Theta$ function recovers its original form after the rescaling operation \cite{RG_mixedsys}. Therefore, the Kondo interaction is irrelevant: 
    $[J]=-1$. 
Since the Kondo interaction has been shown to be irrelevant in the weak-coupling limit, the corresponding Kondo vertex will not be included in the subsequent RG analysis. As a result, the RG equations for the composite system decouple into separate RG equations for the conduction electrons and the Majorana fermions $\chi$ separately. By known results, the quartic interaction of $\chi$ is irrelevant at tree level: 
    $[u_{\chi}]=-1$.

Now we focus on the RG equation of the e-e interactions Eqs.~(\ref{S1}) and (\ref{S2}). By standard RG calculations for fermion \cite{shankar1994renormalization}, it can be found that the relevant interactions can be decomposed into $d$-wave spin singlet pairing and $f$-wave spin triplet pairing components:
\begin{equation}
\begin{aligned}
     \delta S^{(f)}_{\mathrm{e}-\mathrm{e}} &= u_f\sum_{s_z=\pm1,0}\int_{\Lambda} \mathrm dk\mathrm dk^\prime \ \bar\Delta_{s_z}^{f}(k)\Delta^{f}_{s_z}(k^\prime),
    \\ \delta S^{(d)}_{\mathrm{e}-\mathrm{e}} &= u^{+}_d\int_{\Lambda} \mathrm dk\mathrm dk^\prime \bar\Delta_{+}^{d}(k)\Delta_{+}^{d}(k^\prime)+\ (+\rightarrow-),
\end{aligned}
\end{equation}
where $u_f=\frac{(V_1(\Lambda)-V_2(\Lambda))\cdot J^{2}}{8K}>0$ and $u^{\pm}_d=\frac{3V_1(\Lambda)\cdot J^2}{16K}>0$. $\Delta^f_{s_z}(k)$ denotes the $f$-wave spin triplet pairing order parameter carrying $z$ direction spin $s_z$. For example, $\Delta_{s_z=1}^s(k)= \psi^{(+)}_{k,\uparrow}\psi^{(-)}_{-k,\uparrow}$.  The 
$d\pm id$ spin-singlet pairing order parameters are given by $\Delta^d_{\pm}(k)=\frac{1}{\sqrt2}e^{\pm i\theta_\mathbf k}(\psi^{(+)}_{k,\uparrow}\psi^{(-)}_{-k,\downarrow}-\psi^{(+)}_{k,\downarrow}\psi^{(-)}_{-k,\uparrow})$. The identification of $\Delta^f_{s_z}(k)$ and $\Delta^d_{\pm}(k)$ as $f$-wave and $d$-wave pairing, respectively, can be established by examining the phase winding of $\Delta(\mathbf k)$ around the $\Gamma$ point. The details of the RG analysis are provided in the Supplemental Material. We finally obtain the following RG equations:
\begin{equation}
    \begin{aligned}
    &\frac{\mathrm du_f}{\mathrm dl}=\frac{u^{2}_f}{6\pi t},\\
    &\frac{\mathrm du^{\pm}_d}{\mathrm dl}=\frac{{u^{\pm}_d}^2}{6\pi t},
\end{aligned}
\end{equation}
where $u_f,u^{\pm}_d$ both flow to infinity under RG, indicating that the Fermi surface of the conduction electrons is unstable against the attractive interaction mediated by the exchange a pair of Majorana spinons. Since $u^{\pm}_d/u_f\gg1$, we expect the ground state to host a $d$ wave spin singlet SC order. 

\begin{figure}
\includegraphics[width=\linewidth]{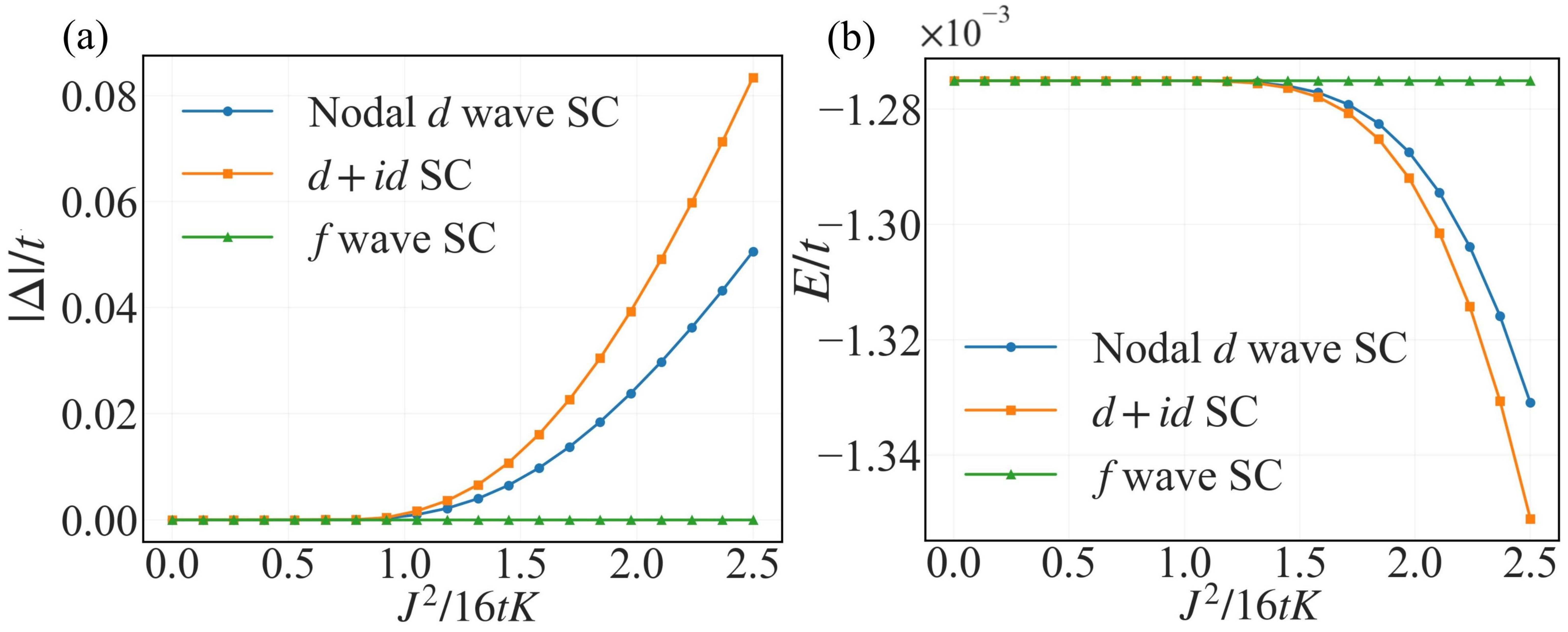 }
\caption{\label{fig: Mean Field results} The figures present the mean-field results for superconductivity of the conduction electrons. Panels (a) and (b) show the order parameters and energies, respectively, as functions of the interaction strength $J^2/16tK$. Results obtained from three different initial conditions in the self-consistent calculations are displayed, corresponding to nodal $d$ wave SC, $d+id$ SC and $f$ wave SC.}
\end{figure}

One way to analyze the possible broken symmetries is to revert to the Hamiltonian formulation and explicitly write out the relevant interactions:
\begin{equation}
  \begin{aligned}
    \hat H_{\mathrm{int}}=-&\int_{\Lambda} \frac{\mathrm d^{2}\mathbf k\ \mathrm d^{2}\mathbf k^{\prime}}{(2\pi)^4}\ u_f \sum_{s_z=\pm1,0}{{\hat{\Delta}^{f\dagger }_{s_z}(\mathbf k)}}\hat\Delta^{f}_{s_z}(\mathbf k^{\prime}) \\
    &+u^{+}_d \hat\Delta^{d\dagger}_{+}(\mathbf k)\hat\Delta^{d}_+(\mathbf k^{\prime}) + \ (+\rightarrow-),
\end{aligned}  
\end{equation}
where the effective interaction strengths $u_f,u^{\pm}_d$ are obtained by integrating out the Majorana spinon $\chi$, which gives $u_f=\frac{V_1(0)-V_2(0)}{8K}J^2\approx0.045J^2/K$ and $u_d=\frac{3V_1(0)}{16K}\cdot J^2\approx0.405{J^2}/{K}$. The effective interactions are imposed on a $2\Lambda$ shell surrounding the conduction electron Fermi surface. We propose the following quadratic Hamiltonian as mean field ansatz:
\begin{equation}
\begin{aligned}
    \hat H_{MF} &= \hat H_0 -\tilde\mu\hat N+ \sum_{s_z=\pm1,0} \Delta^{f}_{s_z}\int_\Lambda  \frac{\mathrm d^2\mathbf k}{(2\pi)^2} \ \hat\Delta^{f}_{s_z}(\mathbf k)\\ &+\Delta^{d}_+\int_{\Lambda} \frac{\mathrm d^2\mathbf k}{(2\pi)^2}\ \hat\Delta^{d}_+(\mathbf k) + \ (+\rightarrow -) + \mathrm {H.c.}.
\end{aligned}
\end{equation}

 The mean field Hamiltonian allows for a mixing of $f$ and $d\pm id$ SC pairing order. The parameters $ \Delta^{f}_{s_z}$ and $ \Delta^{d}_\pm$ characterize the strengths of $f$ and $d\pm id$ pairing order parameters respectively. For given values of $(\Delta^{f}_{s_z}, \Delta^{d}_\pm)$, the chemical potential $\tilde\mu$ is determined self-consistently to ensure particle number conservation. The ground state $\left | \psi_{MF}  \right \rangle$ of $\hat H_{MF}$ can be solved and $(\Delta^f_{s_z},\Delta^{d}_\pm)$ are determined by minimizing the energy $E=\langle \psi_{MF}  | \hat H_0+\hat H_\mathrm {int}|\psi_{MF} \rangle $.

 The basic parameters used in mean field calculation are chosen as follows: $\mu=0.3t$, $\Lambda =\frac{\pi}{10}$. The initial value of $\Delta^{f}_{s_z}$ is always taken to be nonzero. The initial values of $ \Delta^{d}_{\pm}$ are set in three different ways: (1) $\Delta^{d}_{\pm}=0$;(2) $|\Delta^{d}_{+}|>|\Delta^{d}_{-}|>0$; (3)$|\Delta^{d}_{+}|=|\Delta^{d}_{-}|>0$. The (1) condition converges to pure $f$ wave superconductivity. By calculation, it is found that the $f$-wave triplet pairing order parameters converges to two degenerate solutions: (i)$|\Delta^f_{s_z=+1}|=|\Delta^f_{s_z=-1}|=\Delta_0,\  |\Delta^f_{s_z=0}|=0$.\ (ii) $ |\Delta^{f}_{s_z=0}|=\sqrt{2}\Delta_0,\ \Delta^{f}_{s_z=+1}=\Delta^{f}_{s_z=-1}=0$. After self-consistent iterations, we find that the (2) and (3)  converge to $d\pm id$ and nodal $d$ wave superconductivity respectively.  As can be seen in Fig.~$\ref{fig: Mean Field results}$, (a) shows the curves of order parameter in three different conditions as a function of $J^2/16tK$. The $d+id$ SC order opens a larger gap than nodal $d$ wave SC and $f$ wave SC. Fig.~$\ref{fig: Mean Field results}$ (b) presents the energy comparison among different pairing states, indicating that the ground state hosts a pure 
 $d+id$ superconducting order. The chiral $d+id$ SC order spontaneously breaks the time reversal symmetry and exhibits a nontrivial band topology characterized by a Chern number $C=\pm2$.
 
\textit{Conclusion and outlook.}---In this paper, owing to the special $Z_2$ gauge structure of the Yao-Lee-Kondo model that allows us to perform a perturbatively exact RG analysis of the model, we have demonstrated that topological $d+id$ superconductivity emerges in its ground state, regardless of how weak the Kondo coupling is. Moreover, the topological $d+id$ spin singlet SC order co-exists with an underlying fractionalized Yao-Lee spin liquid, realizing an SC* phase. Therefore, the Yao-Lee-Kondo model provides a novel mechanism for the emergence of a fractionalized topological $d+id$ SC phase in correlated systems. We note that the SC* phase in the weak coupling regime shows a very different pairing symmetry from the triplet SC phase in the intermediate coupling regime, where the topological order of the Yao-Lee spin liquid is destroyed. The precise nature of the phase transition between the topological singlet SC* phase and the triplet SC phase remains to be explored in the future. To possibly realize a material platform for the Yao-Lee-Kondo model, building on a microscopic roadmap \cite{Yao-Lee_Roadmap} for realizing the Yao-Lee bond-dependent spin-orbital exchange interaction one may envision heavy-fermion analogs where such a Yao-Lee spin liquid forms in the local-moment sector and is then Kondo-coupled to itinerant conduction electrons.

{\it Acknowledgments}: This work is supported in part by the NSFC under Grant Nos. 12347107 and 12334003 (C.T. and H.Y.), MOSTC under Grant No. 2021YFA1400100 (H.Y.), and the New Cornerstone Science Foundation through the Xplorer Prize (H.Y.).

\bibliography{reference}

\onecolumngrid    
\vspace{1cm}
\begin{center}
    {\bf \Large Supplemental Material}
\end{center}
\setcounter{equation}{0}
\renewcommand{\theequation}{S\arabic{equation}}
\setcounter{section}{0}
\renewcommand{\thesection}{S\arabic{section}}  
\renewcommand{\thefigure}{S\arabic{figure}}    
\renewcommand{\thetable}{S\arabic{table}}      

\subsection{A. The Detailed Derivation of $\delta S^{\mathrm{eff}}_{\mathrm{e-e}}$}

To start with, we write out the action of Kondo interaction after projected to the conduction electron band basis:

\begin{equation} S_{K}=\int_{\Lambda}\mathrm{d}k\mathrm{d}q_1\mathrm{d}q_2\ \sum_{a<b}\frac{J_\alpha^{l,l^\prime}(\theta_\mathbf k,\theta_{\mathbf k^*})}{2}\ \chi^{a}_{q_1,\alpha}\chi^{b}_{q_2,\alpha}\cdot\bar \psi^{(l)}_{k^*}\sigma^{a}\sigma^{b}\psi^{(l^{\prime})}_{k}\ \Theta(\Lambda-|\tilde k^{*}|)\  .
\end{equation}
Here the momentum $\mathbf k$ and $\mathbf k^{*}$ is restricted to the $2\Lambda$ shell surrounding Fermi surface and $\chi^{a}_{q_1,\alpha},\chi^{b}_{q_2,\alpha}$ are the fast modes to be integrated out. $k^*$ is determined by momentum conservation. In the maintext, Eq.~(\ref{S_bare}) shows the $S_K$ in low energy regime and the momentum conservation gives $\delta$ function about valley index $l$. The Kondo coupling amplitude $J^{l,l^\prime}_{\alpha}(\theta_\mathbf k,\theta_{\mathbf k^*})$ takes the form:
\begin{equation}
    J^{l,l^\prime}_{\alpha}(\theta_\mathbf k,\theta_{\mathbf k^*})=
\begin{cases}
\frac{ ll^\prime J}{2}e^{i(l^{\prime}\theta_\mathbf k-l\theta_\mathbf {k^*})}, & \alpha = 1, \\
\frac{J}{2}, & \alpha = 2 .
\end{cases} .
\end{equation}
Here we use the gauge $\psi^{(l)}_{\mathbf k}=\frac{1}{\sqrt 2}(le^{il\theta_\mathbf k},1)^{\mathrm T}$ for band wavefunction. Then we compute the  effective e-e interaction induced by exchanging a pair of Majorana spinons. We focus on the one-loop Feynman diagram:
\begin{figure}[h!]
\includegraphics[width=0.3\linewidth]{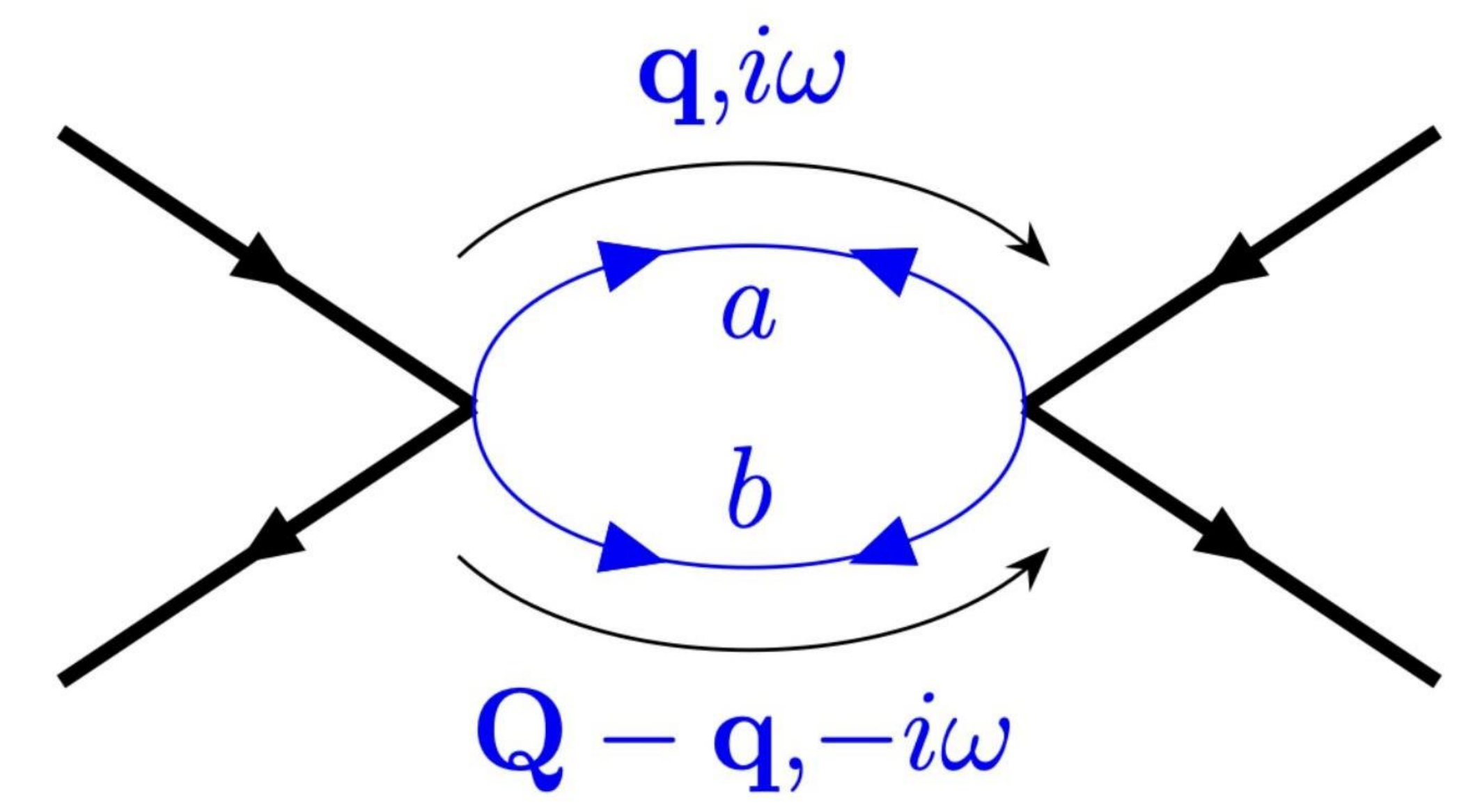}
\end{figure}
\begin{equation}
    \delta S^{\mathrm{eff}}_{\mathrm{e-e}}=\frac{1}{8}\int_\mathrm{\Lambda}\mathrm{d}k_1\mathrm{d}k^{\prime}_1\mathrm{d}k_2\ V_{\alpha\beta}(\Lambda,\mathbf Q)\ J^{l_1,l_1^\prime}_\alpha(\theta_{\mathbf k^{\prime}_1},\theta_{\mathbf k_1}) J^{l_2,l_2^{\prime}}_\beta(\theta_{\mathbf k^{*}_2},\theta_{\mathbf k_2})\ \bar \psi^{(l_1)}_{k^\prime_1}\vec{\sigma} \psi^{(l_1^{\prime})}_{k_1}\cdot \bar \psi^{(l_2)}_{ k^{*}_2}\vec{\sigma}\psi^{(l^{\prime}_2)}_{k_2}\Theta_{\mathbf{k}^{*}_2,\Lambda},
\label{S3}
\end{equation}
where $\Theta_{\mathbf{k}^{*}_2,\Lambda}$ is the step function to constrain $\mathbf k_2$ in the low energy regime surrounding Fermi surface and $\mathbf Q=\mathbf k_1- \mathbf k^{\prime}_1 + \mathrm {valley\ term}$. The frequency dependency is ignored in calculation and $\delta S^{\mathrm{eff}}_{\mathrm{e-e}}$ takes the form of ferromagnetic/antiferromagnetic coupling which depends on the sign of $V_{\alpha\beta}$. $V_{\alpha\beta}(\Lambda,\mathbf Q)$ is defined by the integration of Majorana propagator part in the regime that both $\mathbf K$ and $\mathbf{Q-K}$ are fast modes:
\begin{equation}
    V_{\alpha\beta}(\Lambda,\mathbf Q) = \int_{\mathbf q,\mathbf {Q-q} \notin\Lambda}\frac{\mathrm{d}^{2}\mathbf q}{(2\pi)^2}\mathrm{d}\omega \ G^{\chi}_{0,\alpha\beta}(\mathbf q,\omega)G_{0,\alpha\beta}^{\chi}(\mathbf{Q}-\mathbf{q},-\omega).
\end{equation}
 The Green's function $G^{\chi}_{0,\alpha\beta}$ can be explicitly written out in orbital basis:
\begin{equation}
    G^{\chi}_{0,\alpha\beta}(\mathbf q,\omega)=\sum_n\frac{\chi_n(\mathbf q,\alpha)\chi^{*}_n(\mathbf q,\beta)}{-i\omega+\epsilon_n^{\chi}(\mathbf q)},
\end{equation}
where $n$ is the band index for Majorana. $\chi_n(\mathbf q,\alpha)$ and $\epsilon_n^{\chi}(\mathbf q)$ are the band wavefunction and band dispersion of $\chi$. The flavor index of $\chi$ is neglected here because of $SO(3)$ symmetry. After carrying out the $\omega$ integral:
\begin{equation}
    V_{\alpha\beta}(\Lambda,\mathbf Q)=\sum_n \int_{\mathbf q,\mathbf Q-\mathbf q\notin\Lambda}\frac{\mathrm d^2\mathbf q}{(2\pi)^2}\frac{2\pi\cdot\chi_n(\mathbf q,\alpha)\chi^{*}_{n}(\mathbf q,\beta)\chi_n(\mathbf Q-\mathbf q,\alpha)\chi^{*}_n(\mathbf Q-\mathbf q,\beta)}{|\epsilon^{\chi}_n(\mathbf q)+\epsilon^{\chi}_n(\mathbf Q-\mathbf q)|}.
    \label{V_integral}
\end{equation}
To qualitatively analyze the induced e-e interaction with a small enough Fermi surface, we take the approximation that the momentum transfer $\mathbf Q\approx0 ,\pm2\mathbf K$, corresponding to intra-valley and inter-valley scattering of conduction electron respectively. Take the limit $\Lambda=0$, the matrix $V(0,\mathbf Q)$ is numerically computed out:
\begin{equation}
    V(\Lambda =0,\mathbf{Q}=0)=\frac{V_1}{K}\begin{pmatrix}
1  & -1\\
-1  & 1
\end{pmatrix},\ \ \     V(\Lambda =0,\mathbf Q=\pm2\mathbf K)=\frac{V_2}{K}\begin{pmatrix}
1  & 0\\
0  & 1
\end{pmatrix}
\label{V_matrix},
\end{equation}
where dimensionless constant $V_1\approx 2.17,V_2\approx1.80$. With cutoff $\Lambda$ tuned slightly away from 0, $V_1$ and $V_2$ are substituted by function $V_1(\Lambda)$ and $V_2(\Lambda)$. Therefore, the induced $\mathrm e$-$\mathrm e$ interaction includes intra-orbital ferromagnetic coupling and inter orbital anti-ferromagnetic coupling. 

Here we give a brief explanation about the sign of integral Eq.~(\ref{V_integral}) when $\mathbf Q=0$. The time reversal symmetry of the Majorana free Hamiltonian $\hat h^{(\chi)}$ can be written as $\mathcal T=C\mathcal{K}$. $\mathcal K$ is complex conjugate operator and $C$ is sublattice opeartor with $C_\alpha=\pm 1$ for $\alpha = A/B$ orbital. This gives us the relation: 
\begin{equation}
    \chi_n(\mathbf K,\alpha)=C_{\alpha}\chi^{*}_n(-\mathbf K,\alpha),
\end{equation}
which gives:
\begin{equation}
 \mathrm{Numerator\ of \ Eq.~(\ref{V_integral}})= C_{\alpha}C_{\beta}\cdot |\chi_n(\mathbf K,\alpha)|^2|\chi_n(\mathbf K,\beta)|^2\equiv \frac{C_\alpha C_\beta}{4}.
\end{equation}

By summing over $\alpha,\beta$ in Eq.~(\ref{S3}), we obtain the final results $\delta S^{\mathrm{eff}}_{\mathrm{e-e}}=\delta S^{(1)}_{\mathrm {e-e}}+\delta S^{(2)}_{\mathrm {e-e}}$, where $\delta S^{(1)}_{\mathrm {e-e}}$ and $\delta S^{(2)}_{\mathrm {e-e}}$ correspond to the $\mathbf Q=0$ and $\mathbf Q=\pm 2\mathbf K$ terms respectively:
\begin{equation}
    \delta S^{(1)}_{\mathrm{e}-\mathrm{e}} = \frac{V_1(\Lambda)J^2}{16K}\int_{\Lambda} \mathrm dk\mathrm dk^\prime \ [1-\cos(\theta_\mathbf k-\theta_{\mathbf k^{\prime}})]\ \bar\psi^{(l)}_k\vec{\sigma}\psi^{(l)}_{k^\prime}\cdot\bar\psi^{(\bar l)}_{-k}\vec\sigma\psi^{(\bar l)}_{-k^{\prime}}
    \label{Seff_1}
\end{equation}

\begin{equation}
    \delta S^{(2)}_\mathrm{e-e}=\frac{V_2(\Lambda)J^2}{16K}\int_{\Lambda} \mathrm{d}k\mathrm{d}k^{\prime}\ \bar\psi^{(l)}_k\vec{\sigma}\psi^{(\bar l)}_{k^{\prime}}\cdot\bar\psi^{(\bar l)}_{-k}\vec\sigma\psi^{(l)}_{-k^{\prime}}
    \label{Seff_2}
\end{equation}
Here the interactions have been written in the Cooper channel.

\subsection{B. RG calculation of $e-e$ interaction}
The quartic interaction terms in Cooper channel can be written in the general form: 
\begin{equation}
    S=\sum_{q,q^{\prime},m,n}\bar\Delta_{q,m}\Gamma^{q,q^{\prime}}_{m,n}\Delta_{q^{\prime},n},
\end{equation}
where $\{\Delta_{q,m}\}$ forms a orthogonal and complete basis of the pairing operator between $q$ and  $-q$ electrons. For a compact form of the RG equation, we redefine the modified $\Gamma$ matrix:
\begin{equation}
    \tilde\Gamma_{m,n}^{q,q^{\prime}}=\frac{1}{\sqrt{v_{\mathbf q}v_{\mathbf q^{\prime}}}}\Gamma_{m,n}^{q,q^{\prime}},
\end{equation}
$v_{\mathbf q}=|\nabla_{\mathbf q}\epsilon_\mathbf q|$ denotes the Fermi velocity at $\mathbf q$. In our model, $v_\mathbf q\approx v_F=\frac{3}{2}t$ with a relatively small $\mu$. By standard RG calculation for fermions, the RG equation for $\tilde\Gamma$ can be written as:
\begin{equation}
    \frac{\mathrm{d} \tilde\Gamma}{\mathrm{d} l} =\frac{1}{4\pi}\tilde\Gamma^{2} .
    \label{RG_Gamma}
\end{equation}
By diagonalizing $\tilde\Gamma$, the pairing terms with a positive eigenvalue $\lambda>0$ correspond to relevant attractive interaction that flows to infinite under RG. And the $\lambda=0$ and $\lambda<0$ components correspond to the marginal and irrelevant part.

We can decouple the interaction part into $f$ wave and $d\pm id$ wave part. To consider the $d+id$ wave part first:
\begin{equation}
    \delta S^{(d+id)}_{\mathrm{e}-\mathrm{e}} = -\frac{V_1(\Lambda)J^2}{16K}\int_{\Lambda} \mathrm dk\mathrm dk^\prime \ e^{i(\theta_\mathbf {k^{\prime}}-\theta_{\mathbf k}) }\bar\psi^{(+)}_k\vec{\sigma}\psi^{(+)}_{k^\prime}\cdot\bar\psi^{(-)}_{-k}\vec\sigma\psi^{(-)}_{-k^{\prime}}.
    \label{S d+id}
\end{equation}
 We focus on the $s_z=0$ pairing terms, which contains both singlet and triplet pairing components. We write Eq.~(\ref{S d+id}) in the bylinear form about $\Delta^{(d+id)}_{\sigma}(k)=e^{i\theta_\mathbf k} \psi^{(+)}_{k,\sigma}\psi^{(-)}_{-k,\bar \sigma}$. The corresponding $\Gamma$ matrix can be written as: $\Gamma^{q,q^\prime}_{\sigma,\sigma^{\prime}}= \frac{V_1(\Lambda)}{16K}J^2\cdot (-2\sigma_x+\mathbb I)_{\sigma\sigma^{\prime}}$. Therefore, the $d\pm id$ interaction gives $u_d=\frac{3V_1(\Lambda)}{16}\frac{J^2}{K}$ relevant singlet pairing term with order parameter:
\begin{equation}
    \Delta^d_{\pm}(k)=\frac{1}{\sqrt2}e^{\pm i\theta_\mathbf k}(\psi^{(+)}_{k,\uparrow}\psi^{(-)}_{-k,\downarrow}-\psi^{(+)}_{k,\downarrow}\psi^{(-)}_{-k,\uparrow}).
\end{equation}
Then we consider the $f$ wave interaction and also focus on the $s_z=0$ pairing component. We write out the interactions under the basis $\Delta^{(s)}_{\sigma}(k)= \psi^{(+)}_{k,\sigma}\psi^{(-)}_{-k,\bar \sigma}$. The $\Gamma$ matrix for Eqs.~(\ref{Seff_1})(\ref{Seff_2}) are $\Gamma^{(1)}_{\sigma,\sigma^{\prime}}=\frac{V_1(\Lambda)}{8K}J^2(2\sigma_x-\mathbb I)_{\sigma\sigma^{\prime}}$ and $\Gamma^{(2)}_{\sigma,\sigma^{\prime}}= \frac{V_2(\Lambda)}{8K}J^2(-2\mathbb I+\sigma_x)$ respectively. The $\Gamma$ matrix for $s$ wave can be written as:
 \begin{equation}
     \Gamma^{q,q^\prime}_{\sigma,\sigma^{\prime}}= \frac{J^2}{8K}\cdot [-(V_1+2V_2)\mathbb I+(2V_1+V_2)\sigma_x]_{\sigma\sigma^{\prime}}.
 \end{equation}
 The relevant part gives the $f$ wave triplet pairing $\Delta^{f}_{s_z=0}(k)=\frac{1}{\sqrt2}(\psi^{(+)}_{k,\uparrow}\psi^{(-)}_{-k,\downarrow}+\psi^{(+)}_{k,\downarrow}\psi^{(-)}_{-k,\uparrow})$ and the interaction strength is $u_f=\frac{V_1(\Lambda)-V_2(\Lambda)}{8K}J^2$. By the $SU(2)$ symmetry, the complete triplet pairing interaction can be written out:
 \begin{equation}
      \delta S^{(s)}_{\mathrm{e}-\mathrm{e}} = u_f\sum_{s_z=\pm1,0}\int_{\Lambda} \mathrm dk\mathrm dk^\prime \ \bar\Delta_{s_z}^{f}(k)\Delta^{f}_{s_z}(k^\prime).
 \end{equation}
 The RG equation about $\tilde\Gamma$ Eq.~(\ref{RG_Gamma}) can be written in the eigenvector basis: $\frac{\mathrm{d} \lambda}{\mathrm{d} l} =\frac{1}{4\pi}\lambda^{2}$. Then we can obtain the RG equation for $u_f,u^{\pm}_{d}$:
 \begin{equation}
    \begin{aligned}
    &\frac{\mathrm du_f}{\mathrm dl}=\frac{u^{2}_f}{6\pi t},\\
    &\frac{\mathrm du^{\pm}_d}{\mathrm dl}=\frac{{u^{\pm}_d}^2}{6\pi t}.
\end{aligned}
\end{equation}

\end{document}